\documentclass[12pt]{article}
\usepackage{graphicx}
\usepackage{verbatim} 

\newcommand\pubnumber{DPF2013-54}
\newcommand\pubdate{\today}

\def\OSU{Department of Physics\\
The Ohio State University, 
191 West Woodruff Avenue, Columbus Ohio 43210, USA}
\def\Siegen{Universitat Siegen\\
Fachbereich Physik, D57068
Siegen, Germany}
\def\support{\footnote{Work supported by the Department of Energy and the NSF's MRI program.}}

\def\Title#1{\begin{center} {\Large #1 } \end{center}}
\def\Author#1{\begin{center}{ \sc #1} \end{center}}
\def\Address#1{\begin{center}{ \it #1} \end{center}}

\newcommand\pubblock{\rightline{\begin{tabular}{l} \pubnumber\\
         \pubdate  \end{tabular}}}
\newenvironment{Abstract}{\begin{quotation}  }{\end{quotation}}
\newenvironment{Presented}{\begin{quotation} \begin{center} 
             PRESENTED AT\end{center}\bigskip 
      \begin{center}\begin{large}}{\end{large}\end{center} \end{quotation}}
\def\Acknowledgments{\bigskip  \bigskip \begin{center} \begin{large}
             \bf ACKNOWLEDGMENTS \end{large}\end{center}}

\def\beq{\begin{equation}}
\def\eeq#1{\label{#1}\end{equation}}
\def\eeqn{\end{equation}}

\def\beqa{\begin{eqnarray}}
\def\eeqa#1{\label{#1}\end{eqnarray}}
\def\eeqan{\end{eqnarray}}

\let\bar=\overbar

\def\etal{{\it et al.}}

\def\Dslash{\not{\hbox{\kern-4pt $D$}}}
\def\dslash{\not{\hbox{\kern-2pt $\del$}}}

\def\msb{{\bar{\ssstyle M \kern -1pt S}}}

\begin{document}
\begin{titlepage}
\pubblock

\vfill
\Title{The new radiation-hard optical links for the ATLAS pixel detector}
\vfill
\Author{ Richard Kass\support, K.~K.~Gan, H.~P.~Kagan, J.~Moss, J.~Moore, S. Smith, Y.~Yang}
\Address{\OSU}
\Author{P.~Buchholz, M.~Ziolkowski}
\Address{\Siegen}
\vfill
\begin{Abstract}
The ATLAS detector is currently being upgraded with a new layer of pixel based
charged particle tracking and a new arrangement of the services for the pixel detector.
These upgrades require the replacement of the opto-boards previously used by the pixel detector. 
In this report we give details on the design and production of the new opto-boards.
\end{Abstract}
\vfill
\begin{Presented}
DPF 2013\\
The Meeting of the American Physical Society\\
Division of Particles and Fields\\
Santa Cruz, California, August 13--17, 2013\\
\end{Presented}
\vfill
\end{titlepage}
\def\thefootnote{\fnsymbol{footnote}}
\setcounter{footnote}{0}

\section{Introduction}

\begin{figure}[htb]
\begin{center}
 \includegraphics[height=1.5in]{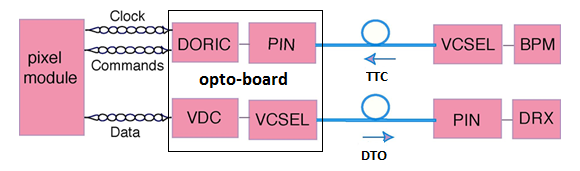}

  {\caption{\label{fig:opto-path}\small 
The routing of signals to and from the opto-board. One link of an opto-board is shown.}}
\end{center}
\end{figure}

The ATLAS experiment is one of two large general purpose spectrometers at CERN's Large Hadron Collider  
(LHC) whose goal is to measure the production and decay characteristics of the Higgs boson(s) and discover
new particles that are not part of the standard model. An important sub-system of the ATLAS detector is
its pixel~\cite{pixel} detector. As originally  built the pixel detector consists of three
concentric tracking layers and three disks on each side of the interaction region. In all, there are
1744 pixel modules with a total of 80 M channels of electronics in the detector. 
Signals  from the control room to a pixel module 
are sent at 40 Mb/s over  80 m of optical fiber to an opto-board~\cite{Arms} and received by a PIN diode. 
An ASIC (DORIC) on the opto-board then decodes
the bi-phase mark encoded signals and transmits the information to a pixel module  using twisted pair
copper cables. Transmission of data from a pixel module to the electronics located outside of the ATLAS
detector is also routed through the opto-board. In this case an ASIC (VDC) on the opto-board 
converts the signals sent over a twisted pair
 of copper wires
to optical signals using a VCSEL\footnote{VCSEL = Vertical Cavity Surface Emitting Laser diode} and transmits them at up to 160 Mb/s. 
The routing of signals to and from the opto-board is shown in Fig~\ref{fig:opto-path}. Each opto-board 
provides 6-8 links, with each link  consisting of a VCSEL/PIN pair that services a pixel module.

While
the performance of the pixel detector was outstanding during the 7 TeV and 8 TeV data runs
it was felt that its future performance and longevity could be enhanced by adding a 4th layer 
of pixel tracking immediately outside the beampipe. This new tracking layer is called
the Insertable B-Layer (IBL). In addition, in order to make the pixel services
more accessible for potential repairs, the location and organization of the services was
re-worked. This project goes by the name of new Service Quarter Panels (nSQP). As a result
of the IBL and nSQP upgrades, the opto-boards used for the 7 TeV and 8 TeV data runs have to be replaced.
In the following section we give the details of these re-designed opto-boards.

\section{The new pixel opto-boards}
\begin{figure}[htb]
\begin{center}
 \includegraphics[height=1.5in]{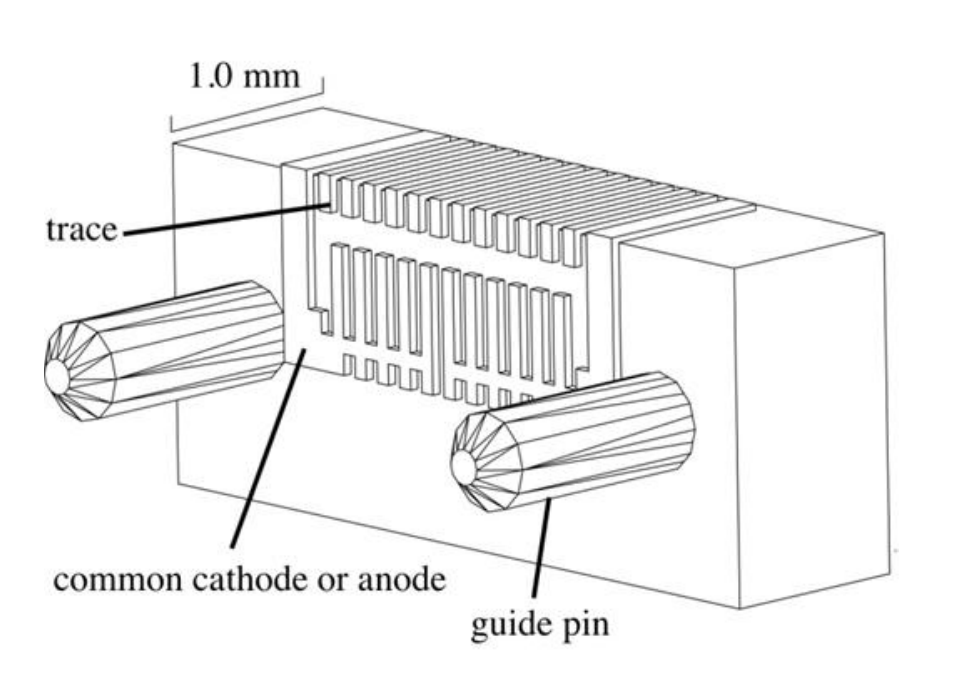}
\includegraphics[height=1.5in]{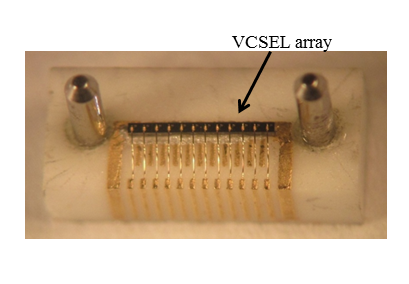}
 \includegraphics[height=1.5in]{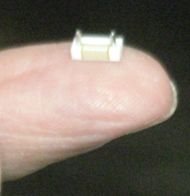}
  {\caption{\label{fig:optopack}\small 
(left) A 3D model of an opto-pack. (middle) A close up photograph of an opto-pack with a wire bonded VCSEL array.
          (right) Scale photograph of an opto-pack.}}
\end{center}
\end{figure}

There are several improvements in the design of the new opto-boards that take advantage of
the experience gained from the production and operation of the first generation opto-boards.
The new design  eases their production and enhances their reliability.
One major improvement is the re-design of the optical packaging of the VCSEL 
and PIN arrays.
The original optical packages (opto-packs) were provided by Taiwan as an
 in-kind contribution to ATLAS. These opto-packs had several drawbacks. First, the VCSEL arrays
were found to fail under moderate humidity and a suitable replacement had to be found.
We measured the radiation hardness~\cite{Gan-Nagarkar} and robustness (e.g. humidity resistance) 
of several commercial VCSEL arrays and chose an array manufactured by Finisar~\cite{Finisar}.
We also decided to switch to a more robust PIN array manufactured by ULM Photonics~\cite{ULM} after measuring
the radiation hardness of several other candidates. In Fig.~\ref{fig:optopack} we show
pictures of the new opto-pack.

\begin{figure}[htb]
\begin{center}
\includegraphics[width=6in]{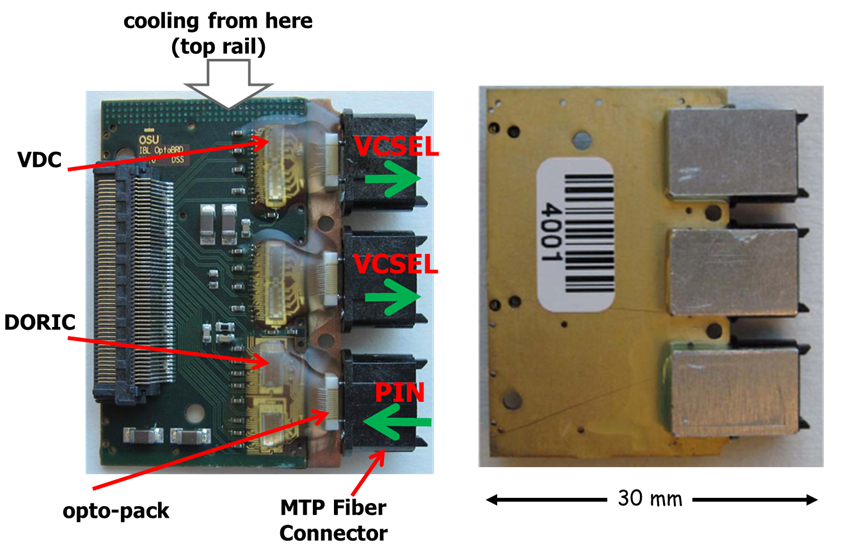}
 {\caption{\label{fig:NewOpto} A photograph of
the re-designed opto-board showing the VCSEL driver chips (VDCs) that couple to the VCSEL opto-packs and
the receiver chips (DORICs) that couple to the PIN opto-pack. Both ASICs were designed by OSU/Siegen
for the first generation opto-boards.}}
\end{center}
\end{figure}

Another drawback of the Taiwan opto-packs had to do with the soldering of the micro-leads on an opto-pack to 
an opto-board. Since the opto-board was fabricated out of BeO to take advantage of its excellent thermal conductivity 
it was very difficult
to provide the appropriate amount of heat to solder a small
lead to the BeO substrate. In fact, it has been determined from the extracted opto-board postmortem 
that the major failure mode  was due to cold solder joints. To avoid this
pitfall, the new opto-boards rely on wire bonds rather than solder~\cite{Gan}.
Other improvements include the use of a polyimide PCB mounted to a  copper 
plate instead of a thick film circuit on BeO.  While BeO has
excellent thermal properties it is an expensive material as well as one that requires special
handling due to safety requirements. Our new PCB is a single sided board (the original opto-boards were double sided) and
has a 1 mm thick layer of copper on the backside for thermal management. We also added redundant 
control lines by taking advantage of unused pins on the opto-board's connector as well as
eliminating the daisy chaining of control signals on the opto-board.   In Fig.~\ref{fig:NewOpto} we show
an example of a recently assembled opto-board.

These new opto-boards come in three varieties; IBL, B (originally used by the first layer of the pixel detector),
and D (originally used in the disks and second and third layers of the pixel detector). All three varieties were
extensively proto-typed both in system wide tests and irradiations. Twenty two proto-type boards were
distributed to our collaborators and subjected to a further round of testing. The results of these tests
gave us the confidence that we could move into the production phase.   

\begin{figure}[htb]
\begin{center}
\includegraphics[width=6in]{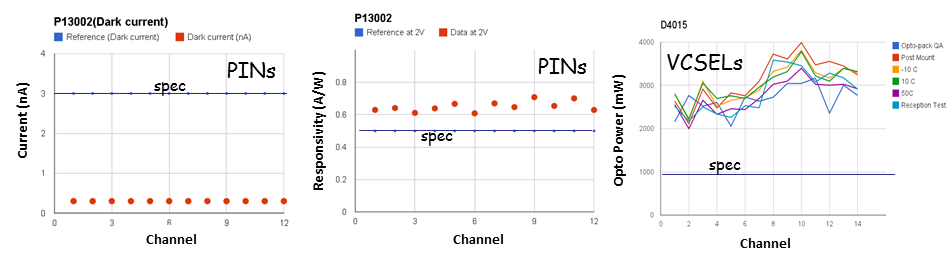}
 {\caption{\label{fig:opto-QA1} (left) Dark current vs. channel number and 
(middle) Responsivity vs. channel number for a typical PIN. (right) Optical power for
a typical VCSEL vs. channel number
at various production stages and test conditions. }}
\end{center}
\end{figure}

The production of the opto-boards (60 IBL, 55 B, 275 D) began in the summer of 2013. 
 This includes
fabricating the opto-packs (1200), mounting of opto-packs and ASICs on the PCB, and wire bonding of 
 opto-packs and opto-boards. Some of this work requires the components (e.g. VCSELs) to be placed with a precision
better than 10 $\mu$m otherwise the optical coupling 
to the fiber will be poor. The two ASICs (VDC, DORIC) on the opto-board
are the same versions used for the previous opto-boards and were designed by OSU and Siegen~\cite{Arms}.  Once an opto-board
is fabricated it is subjected to an extensive array of tests including burn in at 50$^\circ$C, thermal cycling,
measurement of its optical power, rise/fall time, duty cycle, jitter, and bit error rate. Examples
of the results of the quality assurance tests for a typical PIN and VCSEL are shown in Fig.~\ref{fig:opto-QA1}.
Opto-boards that pass all
the quality assurance tests are then shipped to CERN where they are retested 
to insure that there was no damage due to shipping.

\section{Summary}
In this report we have briefly described the assembly and testing of the new opto-boards
being produced for the IBL and nSQP upgrades for the ATLAS detector. These opto-boards are 
expected to be more robust than the previous version of the opto-boards and be used until
the LHC's Long Shutdown 3 (LS3) in the early 2020's.

\Acknowledgments
We have benefited greatly from the support of The Ohio State University physics department, the
Department of Energy, and the NSF. In particular, we wish to acknowledge that this project
could not have succeeded without support from two NSF MRI awards.


\begin{thebibliography}{99}

\bibitem{pixel} G. Aad \etal,
        ``ATLAS Pixel Detector Electronics and Sensors,''
        JINST {\bf 3}, P07007 (2008).

\bibitem{Arms} K.~E. Arms \etal, ``ATLAS Pixel Opto-Electronics,''
         Nucl. Instr. Meth. A {\bf 554}, 458 (2005).
\bibitem{Gan-Nagarkar}  K.K.~Gan \etal, ``Study of the Radiation-Hardness of VCSEL/PIN,''
in Proceedings of the 9th International Conference on Large Scale Applications and Radiation Hardness of Semiconductor Detectors,
Florence, Italy, 2009, PoS (RD09) 041 (2009); A. Nagarkar \etal, ``Study of the Radiation-Hardness of VCSEL \& PIN Diodes,''
in Proceedings of the 11th International Conference on Large Scale Applications and Radiation Hardness of Semiconductor Detectors,
Florence, Italy, 2011, PoS (RD11) 036 (2011).

\bibitem{Finisar} The VCSEL array used is V850-2093-001, fabricated by Finisar, Inc.

\bibitem{ULM} The PIN array used is ULMPIN-04-TN-U0112U, fabricated by ULM Photonics.

\bibitem{Gan} K.K. Gan, 
        ``An MT-Style Optical Package for VCSEL and PIN Arrays,'' 
        Nucl. Instr. and Meth. A {\bf 607}, 527 (2009).







\end{thebibliography}
\end{document}